\documentclass[a4paper]{jpconf}
\usepackage{graphicx}
\begin{document}
\title{Recent Top Properties Measurements at CDF}

\author{Giorgio Chiarelli}

\address{INFN Sezione di Pisa, Largo Bruno Pontecorvo 3, I-56127 Pisa, 
Italy}

\ead{giorgio.chiarelli@pi.infn.it}

\begin{abstract}
We present the most recent CDF results in the measurements of the decay 
and production vertex of the top-quark.
New results on forward-backward asymmetry in top-antitop events are 
presented. Also, recent measurements of the branching fractions of 
top-quark are discussed.
Finally, measurements in single top events, where top-quark is produced 
through electroweak processes, are presented. Despite the much larger 
number of top events collected at the LHC, due to the symmetric initial 
state and the better signal-to-background ratio in specific channels, 
some results will be lasting heritage of the Tevatron. 

\end{abstract}

\section{Introduction}
Top quark was discovered at the Tevatron in 1994-1995. Since then its 
properties were extensively studied by the CDF and D0 experiments.
The last Tevatron run (2001-2011) provided almost 10 fb$^{-1}$ of data to 
each experiment. In this document I will concentrate on the most 
recent CDF results that exploit this wealth of data.

Thanks to its large mass, top is the only quark we can study before 
hadronization, therefore we have the unique opportunity to 
study a "bare" quark. From an experimental point of view, observables are 
(in most cases) perfectly well defined. 
The importance of this very peculiar quark cannot be underestimated. Its 
mass is strictly related, through loops, to the Higgs mass and to vacuum 
stability; its large Yukawa coupling is puzzingly close to one.

Top quark can be produced via electroweak mechanism (single top processes) 
or via strong interactions (in top-antitop pairs). Despite the very small 
signal/background ratio 
for single top production process, both Tevatron experiments detected (and 
studied) top quark in both cases. As top decays in a $W$ and a quark 
(almost 100\% of cases a $b$-quark), 
topologies of top-quark events are classified according to the charged 
boson decay. For strongly produced top events we call "dilepton" events in 
which both $W$s decay leptonically, "l+jets" events in which one of the 
two $W$s decays hadronically and "all-hadron" the case in which both 
bosons decay in quarks{\footnote{In the following I will not present 
results related to the all-hadronic topology.}.
Therefore the typical physics objects present in a top-candidate event of 
interest 
are one or two charged leptons ($e$ or $\mu$), from 2 to 4 jets (two 
coming from $b$-quark hadronization), and large missing transverse energy 
signalling the presence of one or two neutrinos.

In the following analyses we use events with large ($>20$ GeV) $E_T$ 
electrons or muons, large ($>25$ GeV) missing transverse energy (MET) and  
jets of hadrons (reconstructed by a fixed cone algorithm) with $E_T>20$ 
GeV and pseudorapidity $|\eta|<2.8$. 


\section{Study of top-quark properties in $t\bar{t}$ events}
Top production vertex can be studied in top-antitop pair events and in 
events in which top-quark is produced singly. 
CDF studied all topologies and first measured production cross section. 
A summary of the Tevatron results for the top-pair production channel is 
shown in Figure~\ref{xsec}. 
The final CDF result for top-pair production cross section is $\sigma=7.63 
\pm 0.5$ pb, to be compared with the theoretical prediction of 
$\sigma=7.35^{+0.11}_{-0.21}(scale)^{+0.17}_{-0.12}(PDF)$ pb.
The Tevatron result is $\sigma=7.6\pm 0.41(stat)^{+0.2}_{-0.36}(syst.)$, 
and CDF contributes $\approx$ 60 \% to the combination.
\begin{figure}[ht]
\begin{minipage}{2.5in}
\includegraphics[height=2in]{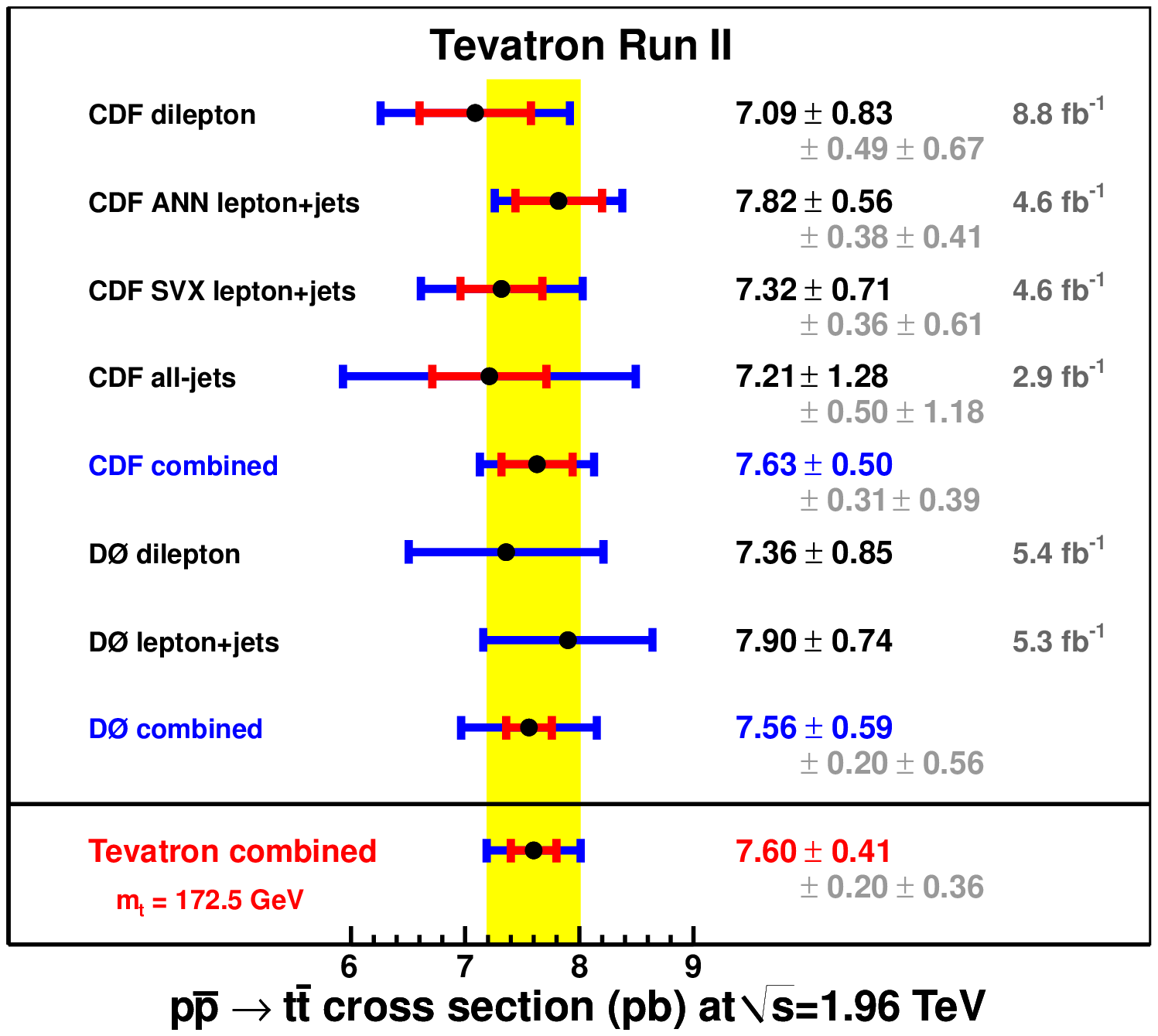}
\caption{\label{xsec}$\sigma_{t\bar{t}}$ measurements at CDF }
\end{minipage}\hspace{2pc}%
\begin{minipage}{2.5in}
\includegraphics[height=2in]{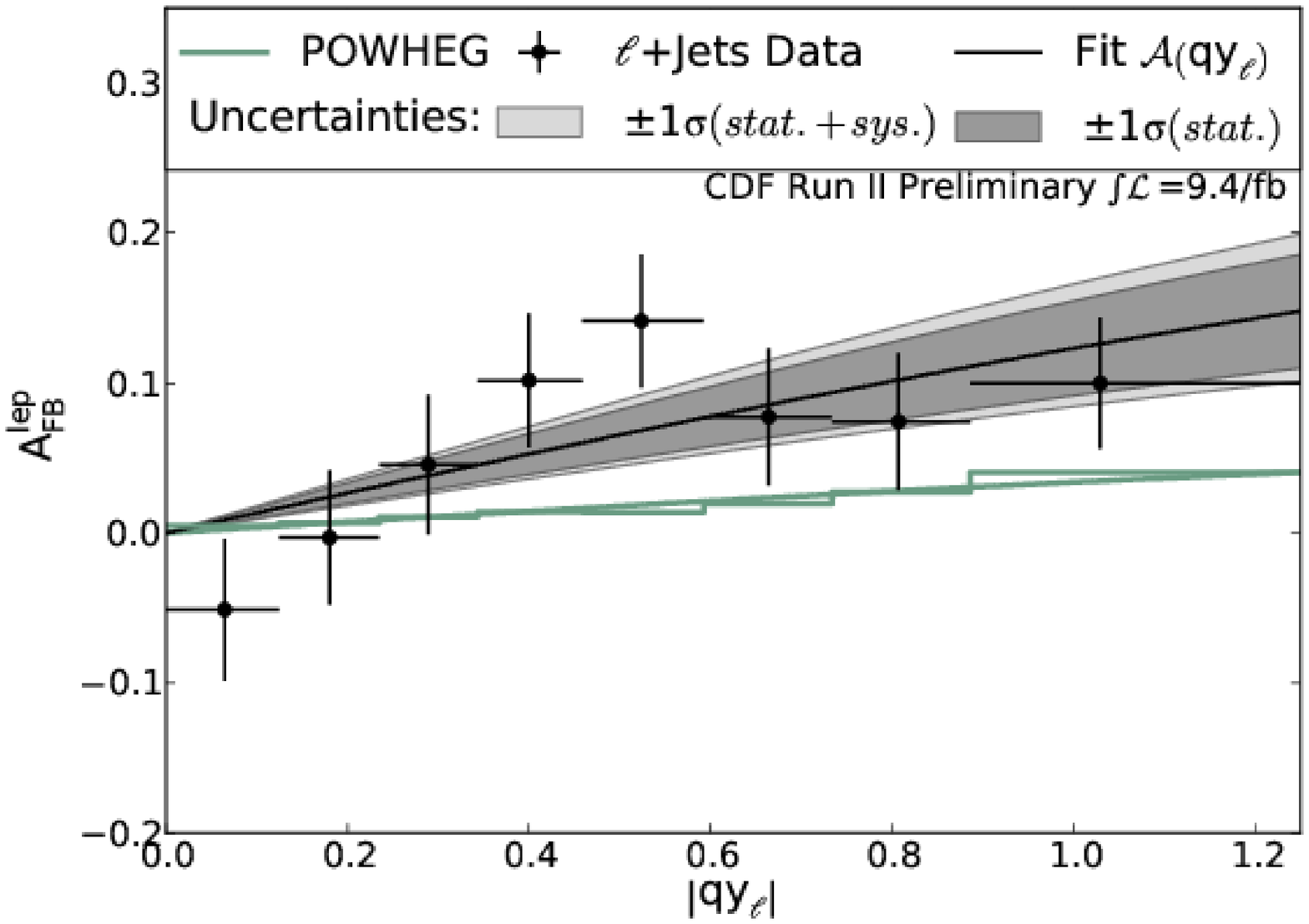}
\caption{\label{afba} Fit to $A_{FB}$ in $l+$jets data.}
\end{minipage}
\end{figure}

\subsection{Angular distributions in $t\bar{t}$ events}
CDF studied the distribution of the angle $\theta$ between the proton and 
the top direction in the $t\bar{t}$ reference frame looking for deviation 
from SM expectations. A description of $d\sigma/d\theta$ in terms of 
Legendre Polynomials shows that the first moment has some tension with the 
SM, but there is not enough sensitivity to the other terms of the 
expansion to draw any conclusion.

Another important distribution to look at is the forward-backward 
asymmetry. Defined as 
$A_{FB}=(N_{\Delta Y>0}-N_{\Delta Y <0})/
(N_{\Delta Y>0}+N_{\Delta Y <0})$, where $\Delta Y$ is the difference in 
rapidity between top and antitop, this quantity is sensitive to new 
physics. 
Positive asymmetry is present in the production process which proceeds via 
$q\bar{q}$ annihilation at tree level or via a box diagram. 
Initial (ISR) and Final (FSR) state radiation can induce a 
negative asymmetry. 
Past CDF observation showed a tension with the SM with $A_{FB}=0.066 \pm 
0.02$ in the $l+jets$ channel. This value, brought back at the parton 
level, translates into an asymmetry of 16.4 $\pm$ 4.7 \%, almost doubling 
the SM prediction~\cite{afb}.

This result prompted a number of studies. Here we present new 
measurements in the $l+$ jets and the dilepton channels and their 
combination.

The asymmetry in the $l+jets$ channel can also be studied in terms of 
direct observables, defining the leptonic asymmetry:   
$A^l_{FB}=(N_{q_l \eta_l>0}-N_{q_l \eta_l <0})/
(N_{q_l \eta_l>0}+N_{q_l \eta_l>0})$, where $l$ indicates the lepton. The 
final result is $A^l_{FB}=9.32^{+3.2}_{-2.9}$ \% against a SM expectation 
of $3.8 \pm 0.3$ \%. Figure~\ref{afba} shows the $A^l_{FB}$ as a function 
of $\eta$~\cite{afbl}.
\begin{figure}[ht]
\begin{minipage}{2in}
\includegraphics[width=2in]{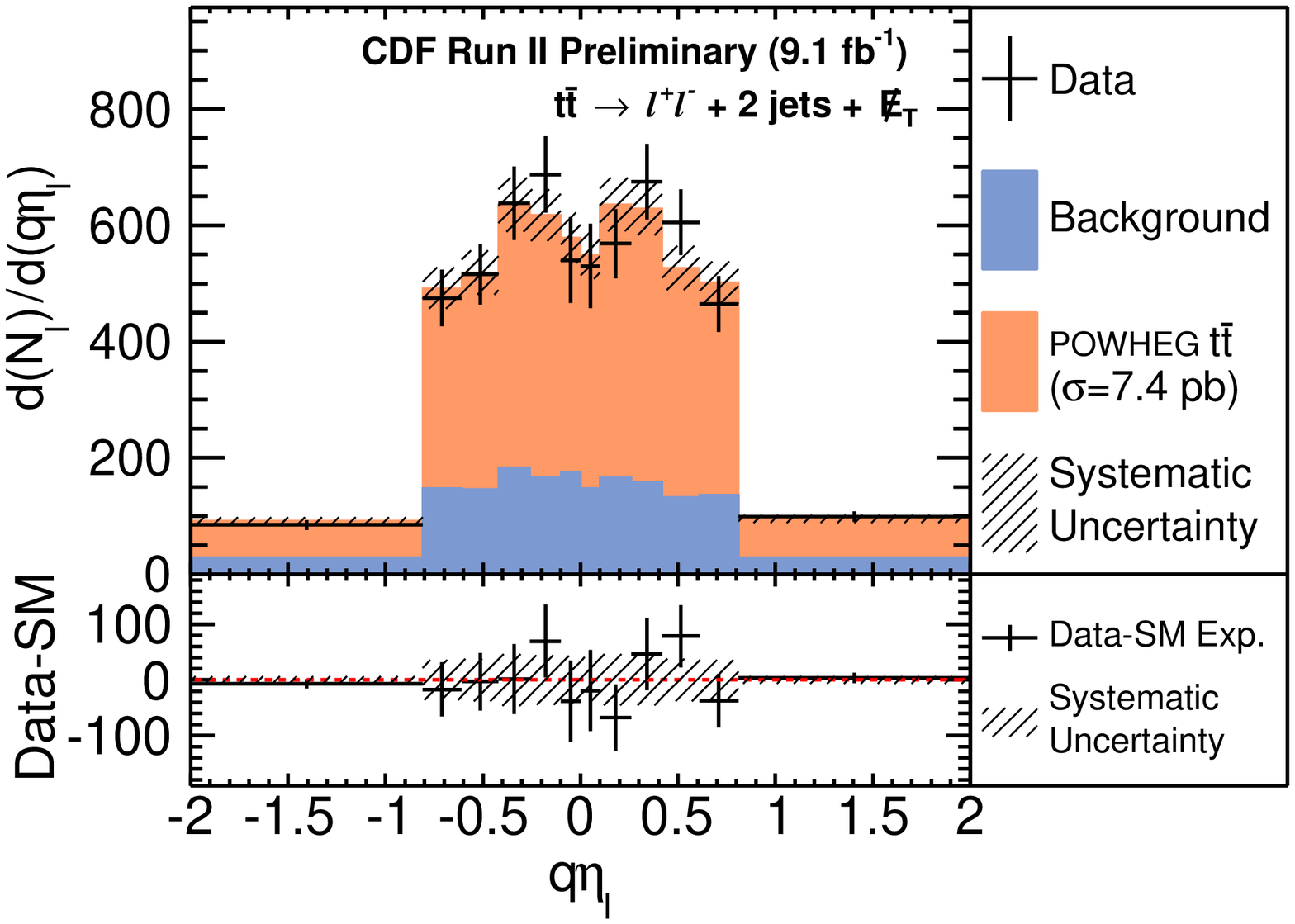}
\caption{\label{Qeta}$q\times \eta$ distribution in dilepton events.}
\end{minipage}\hspace{2pc}%
\begin{minipage}{2in}
\includegraphics[width=2in]{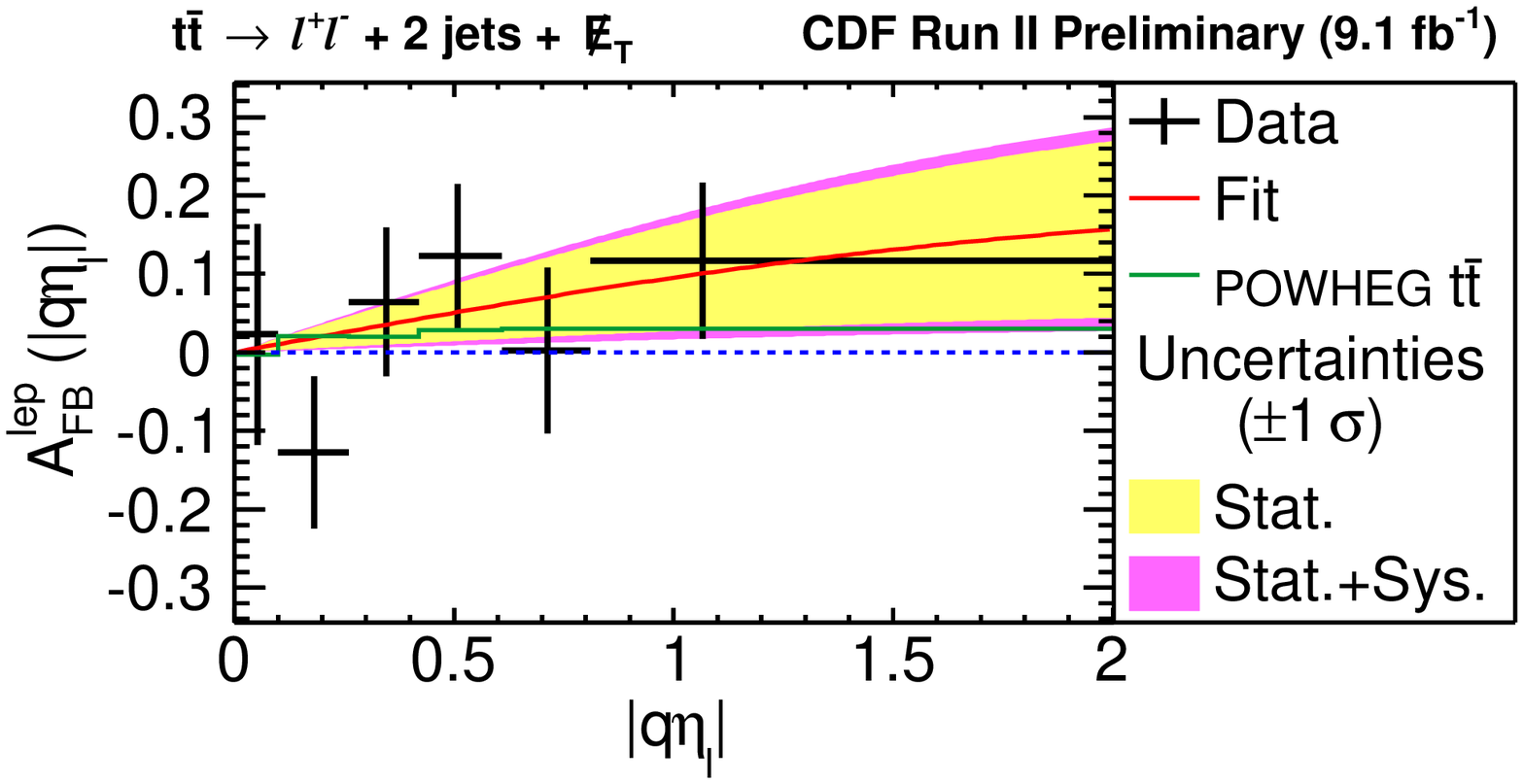}
\caption{\label{asy} Asymmetric part of the $q\times \eta$
distribution compared to different theory predictions.}
\end{minipage}
\end{figure} 
This result shows again tension with the SM.

Recently CDF studied the leptonic asymmetry in the dilepton channel. This 
time one can separately study the symmetric and asymmetric part of the 
distribution of $q\times \eta$  and compare both to 
SM expectations. Figure~\ref{Qeta} shows the distribution of 
$q\times \eta$ compared to 
POWHEG predictions, and Figure~\ref{asy} the asymmetric part. The result,
$A^l_{FB}=(7.2 \pm 5.2(stat) \pm 3(syst))=(7.2 \pm 6.0)$ \% is consistent 
with SM expectation of $(3.8 \pm 3)$ \%. 
In order to improve the experiment sensitivity
one can use both measurements to obtain the best 
estimate. Using 3864 events from the $l+jets$ channel (73\% purity) and 
569 from the 
dilepton channel ($72$\% purity) , CDF combines (using BLUE) an 
asymmetry 
$A_{FB}=(9.0^{+2.8}_{-2.6})$ \%. 
Figure~\ref{alla} 
shows the CDF results compared to SM prediction. The final 
(combined) result is $\simeq 1.8 \sigma$ away by the SM central value.\

\begin{figure}[h]
\begin{minipage}{2.5in}
\includegraphics[height=2in]{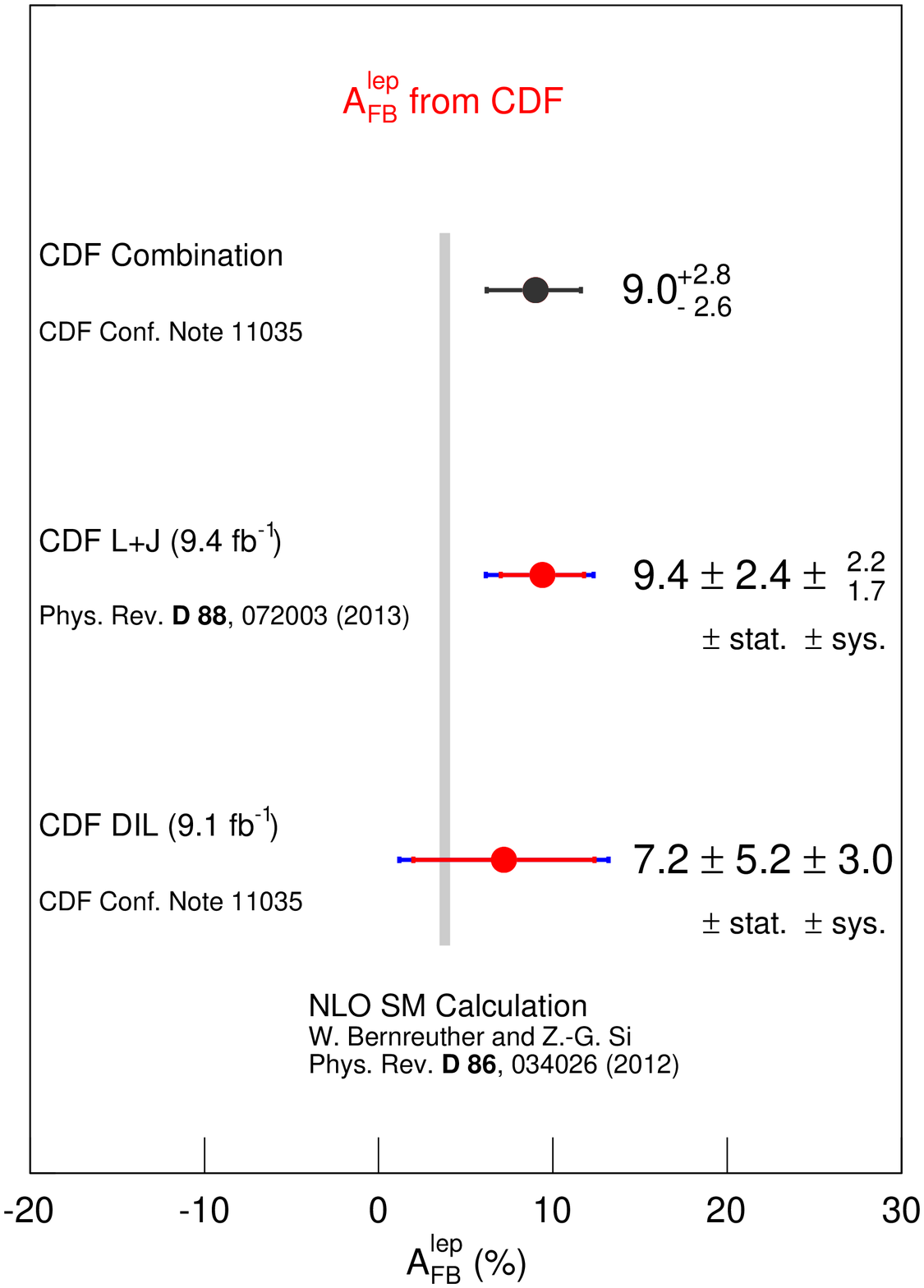}
\caption{\label{alla}Summary of CDF results for $A_{FB}$}
\end{minipage}\hspace{2pc}%
\begin{minipage}{2.5in}
\includegraphics[height=2in]{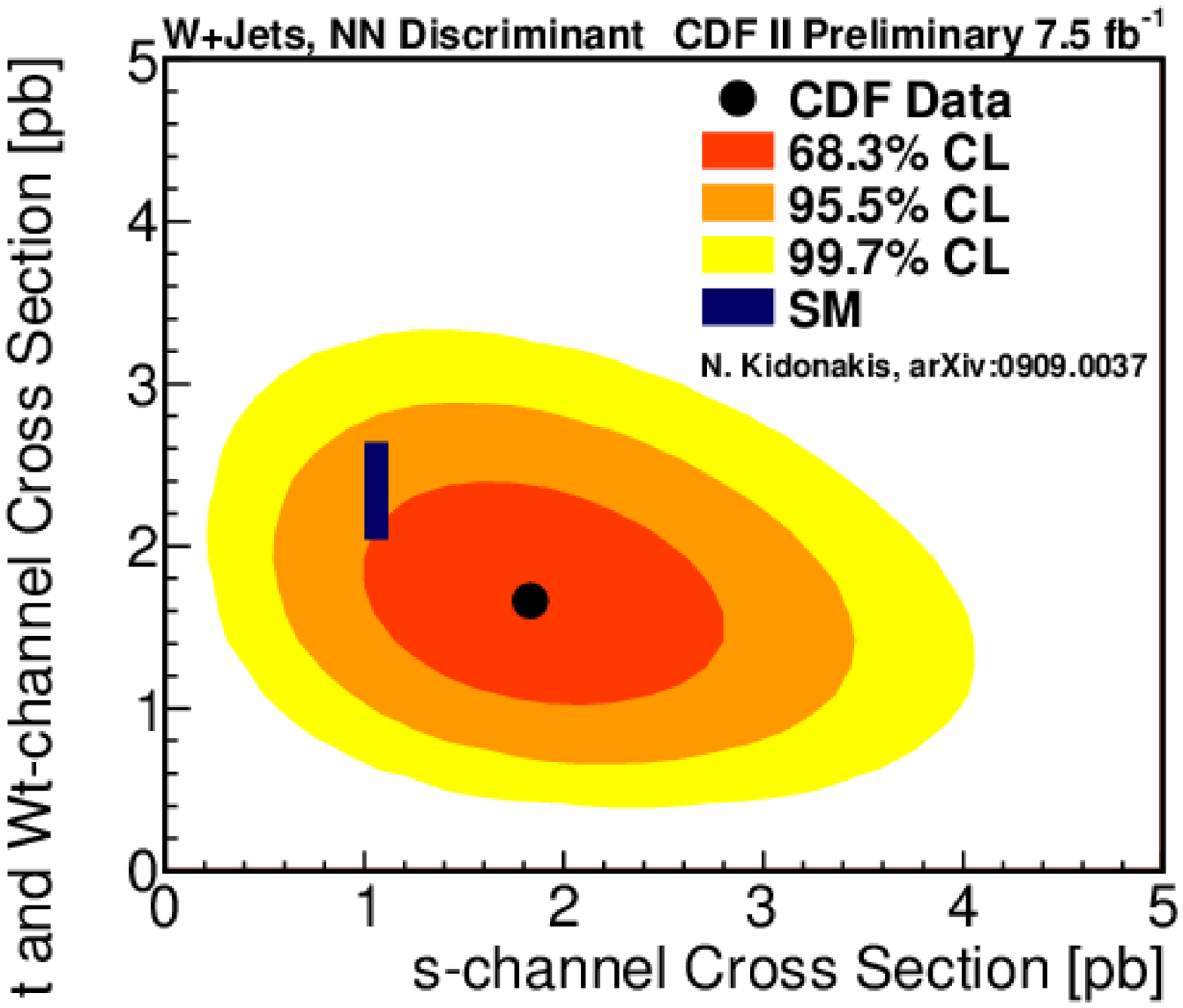}
\caption{\label{spt}Simultaneous $s$ and $t$ cross section determination.}
\end{minipage}
\end{figure}

\subsection{Indirect determination of the CKM element $|V_{tb}|$}
The top-quark decay vertex is related to CKM matrix element $V_{tb}$. One 
can directly measure the branching fractions and there 
are two ways in which one can study $V_{tb}$: indirectly, via ratio 
of branching fractions, or directly, studying single top events.

In the SM, $\sum (|V_{td}|^2+|V_{ts}|^2+(|V_{tb}|^2)=1$, therefore a 
measurement of the ratio of branching fractions 
$R={{|V_{tb}|^2}\over{|V_{td}|^2+|V_{ts}|^2+|V_{tb}|^2}}$ provides an 
estimate of $|V_{tb}|$.

From an experimental point of view, one proceeds by counting the number of 
top candidate events with 0,1,2 b-tagged jets, and comparing these figures 
with expectations as a function of ${R}$. Both CDF and D0 performed 
several measurements in this way. 
The most recent ones are due to CDF in the $l+jets$ and in the 
dilepton channels. In the $l+$jets by 
simultaneously fitting ${R}$ and the $t\bar{t}$ production cross 
section, it measures ${R}=0.94\pm 0.09$ and $\sigma_{t\bar{t}}=7.5\pm 
1.0$ pb, setting a 95 \% C.L. limit of $|V_{tb}|>0.89$~\cite{rl}.

The strategy to extract $R$ in the dilepton sample is different from 
the $l+$jets case 
as, thanks to the very good S/B ratio, it is possible to
first extract the cross section, without using 
$b$-tagging techniques and then measure the b-content of the jets present 
in the events.
Using an integrated luminosity of 8.7 fb$^{-1}$ CDF measures 
${R}=0.87\pm 0.07 (stat+syst)$~\cite{rll} from 
which one extracts $|V_{tb}|=0.93\pm 0.04$ or $|V_{tb}|>0.85$ at 95 \% 
C.L. limit
This result, together with the the D0 result~\cite{d0r} in the dilepton 
channel 
(${R}= 0.86 \pm 0.05$), is tantalizing away from the SM expectation.

\section{Study of top properties in single top events}
Single top production, is the generic definition that encompasses 
electroweak-produced events with only one top (or antitop) quark in the 
final state. 
Besides production through $t$ or $s$ channel (relevant at the Tevatron), 
single top-quark can also be produced in association with a $W$ boson.
Latter process is predicted to be negligible at the Tevatron, and evidence 
for its existence was recently obtained by CMS at the LHC.

Single top production is important, as its cross section is {\it directly} 
proportional to $|V_{tb}|^2$. Therefore it provides access to the $W-t-b$ 
vertex, probes the $V-A$ structure of the theory, gives access to the top 
spin and, of course, makes possible to directly measure $|V_{tb}|$ and 
explore possibilities of new physics.
Besides, existence of FCNC currents or of heavy bosons, Kaluza Klein 
excitations, charged Higgs, may affect differently $t$ and $s$ channel.
Although precision electroweak measurements rule out a fourth generation, 
there is still room for extensions~\cite{jal}.
Precise comparison of 
measurements of single top event properties, could open windows on New 
Physics (if any).

Single top production channels have final state topologies than can be 
mimicked by processes with much larger cross section. Indeed at the 
Tevatron, we are dealing with processes whose rate is smaller than the 
background fluctuations. 
In general, given a topology, we combine several channels, than we develop 
one (or more) Artificial Neural Network(s) (ANN) to identify the signal by 
fitting the appropriate distribution. 
CKM element is extracted by using the relation: $|V_{tb}|^2= 
|V^{SM}_{tb}|^2\times
{\sigma^{obs}\over{\sigma^{SM}}}.$

CDF recently submitted to PRL its result for the $s+t$ 
channel. Using 7.5 fb$^{-1}$ we observe $\sigma_{s+t}=3.04 \pm 
0.55$ pb 
from which we extract $|V_{tb}|=0.95\pm 0.09 (stat.+syst.)\pm 0.05(th.)$, 
and set a limit $|V_{tb}|>0.78$ at 95 \% C.L.
Then, by simultaneously fitting $s$ and $t$ 
components\footnote{$Wt$ component is included in the $t$ channel as it shares the same final 
state topology.}, we extract $\sigma_{t+Wt}=1.66^{+0.53}_{-0.47}$ pb and 
$\sigma_{s}=1.81^{+0.63}_{-0.58}$ pb (see Figure~\ref{spt}), to be 
compared with
SM prediction of $\sigma_{t+Wt}=2.34 \pm 0.3$ pb, and $\sigma_{s}=1.06\pm 
0.06$ pb~\cite{stl}.
If we combine this result with the previous analysis in the MET$+b\bar{b}$ 
channel, we set a limit $|V_{tb}|>0.84$ at 95 \% C.L.

Finally, as the $s-$channel is very difficult to observe at the LHC, both 
CDF and D0 optimized their selections to observe this mode of production.
After this optimization, CDF re-analyzed its MET$+b\bar{b}$ and $l+$jets 
sample, also using a new $b$-tagger developed for the Higgs search.
We find evidence for $s$-channel production at 4.2 $\sigma$, measuring 
$\sigma_{s}=1.38^{+0.38}_{-0.37}$ pb~\cite{cst}.
By combining this result with D0 analysis, the Tevatron experiments 
reported 
the observation of the $s$ channel production at 6.1 $\sigma$, measuring 
$\sigma_{s}=1.29^{+0.26}_{-0.22}$ pb~\cite{tst}.

\subsection{Acknowledgments}
I would like to thank my colleagues D.~Toback, 
J.~Wilson,S.~Leone, and Y.~Peters.

\end{document}